\documentclass[fleqn,usenatbib,useAMS]{mnras}

\usepackage[T1]{fontenc}

\DeclareRobustCommand{\VAN}[3]{#2}
\let\VANthebibliography\thebibliography
\def\thebibliography{\DeclareRobustCommand{\VAN}[3]{##3}\VANthebibliography}

\usepackage{graphicx}	
\usepackage{amsmath}	
\usepackage{orcidlink}
\defcitealias{GQ21}{GQ21}

\usepackage{newtxtext,newtxmath}

\title[Black widow dynamos]{Convective dynamos of black widow companions}

\author[J. Conrad-Burton et al.]{
Jordan Conrad-Burton, Alon Shabi
and Sivan Ginzburg$^{\orcidlink{0000-0002-3751-4553}}$\thanks{E-mail: sivan.ginzburg@mail.huji.ac.il}
\\
Racah Institute of Physics, The Hebrew University, Jerusalem 91904, Israel
}

\date{Accepted XXX. Received YYY; in original form ZZZ}

\pubyear{2023}

\begin{document}
\label{firstpage}
\pagerange{\pageref{firstpage}--\pageref{lastpage}}
\maketitle

\begin{abstract}
Black widows and redbacks are binary millisecond pulsars with close low-mass companions that are irradiated and gradually ablated by the pulsar's high-energy luminosity $L_{\rm irr}$. These binaries evolve primarily through magnetic braking, which extracts orbital angular momentum and pushes the companion to overflow its Roche lobe. Here, we use the stellar evolution code \textsc{mesa} to examine how the irradiation modifies the companion's structure. Strong $L_{\rm irr}$ inhibits convection to the extent that otherwise fully convective stars become almost fully radiative. By computing the convective velocities and assuming a dynamo mechanism, we find that the thin convective envelopes of such strongly irradiated companions ($L_{\rm irr}\gtrsim 3\,{\rm L}_\odot$) generate much weaker magnetic fields than previously thought -- halting binary evolution. 
With our improved magnetic braking model, we explain most observed black widow and redback companions as remnants of main-sequence stars. We also apply our model (with $L_{\rm irr}$) to evolved companions that overflow their Roche lobe close to the end of their main-sequence phase. The evolutionary tracks of such companions bifurcate, explaining the shortest period systems (which are potential gravitational wave sources) as well as the longest period ones (which are the progenitors of common pulsar--white dwarf binaries). The variety of black widow structures and evolutionary trajectories may be utilized to calibrate the dependence of magnetic braking on the size of the convective layer and on the existence of a radiative--convective boundary, with implications for single stars as well as other binaries, such as cataclysmic variables and AM Canum Venaticorum stars.
\end{abstract}




\section{Introduction}

Timing the radio pulse received from millisecond pulsars can reveal low-mass companions around these neutron stars, including the first detected exoplanets \citep{WolszczanFrail92}. Specifically, about a decade ago, dedicated radio surveys and follow-up of {\it Fermi} $\gamma$-ray sources have uncovered a population of binary millisecond `spider' pulsars, whose radio signals are periodically eclipsed by low-mass companions on orbits of several hours \citep{Keith2010, Bates2011, Ray2012, Roberts2013}. These binary pulsars are often divided into `black widows' ($m\approx 0.01-0.05\,{\rm M}_\odot$) and `redbacks' ($m\gtrsim 0.1\,{\rm M}_\odot$), depending on the mass of the companion $m$ (${\rm M}_\odot$ is the solar mass).

Shortly after the discovery of the original black widow \citep{Fruchter1988}, gradual ablation of the companion star by the pulsar's high-energy radiation was invoked to explain both the low mass of the companion and an evaporative wind that causes the radio eclipses (\citealt{Kluzniak1988,Phinney88}; cf. \citealt{EichlerLevinson88,LevinsonEichler91}). Over the last decade, the discovery of a larger population of similar objects motivated a more comprehensive examination of the evolutionary tracks that can transform an initially main-sequence star into a low-mass remnant orbiting a black widow or redback pulsar \citep{Benvenuto2012,Benvenuto2014,Benvenuto2015,Chen2013,JiaLi2015,JiaLi2016,LiuLi2017,Ablimit2019,DeVito2020}. A necessary ingredient in these models is the loss of orbital angular momentum by magnetic braking, which pushes the companion star to overflow its Roche lobe and lose mass (gravitational waves are also efficient at the shortest orbital periods). The magnetic field of low-mass main-sequence stars was historically thought to be generated (at least partially) at the `tachocline' -- the boundary separating the radiative interior from the convective envelope -- commonly leading modellers to turn off magnetic braking once $m\lesssim 0.3\,{\rm M}_\odot$, when stars become fully convective on the main sequence \citep{Rappaport1983,SpruitRitter83}.
Subsequent mass loss is possible through ablation by the pulsar's radiation, which is often parametrized by an ad-hoc efficiency factor linking the pulsar's spin-down power to the ablated wind strength \citep{Stevens1992}.     

\cite{GinzburgQuataert2020}, on the other hand, calculated the strength of the ablated wind from first principles by adapting the \cite{Begelman83} hydrodynamic theory of Compton heated winds \citep[see][for an earlier estimate]{Ruderman89}. The calculated evaporation efficiency appears to be too low for most pulsars to significantly reduce the mass of their companions by ablation directly. Instead, \citet{GinzburgQuataert2020} proposed that the ablated wind couples to the companion's magnetic field, removes angular momentum, and thus maintains stable Roche-lobe overflow. 

\citet[][hereafter \citetalias{GQ21}]{GQ21} demonstrated that this ablation-driven magnetic braking operates at a different rate than standard braking prescriptions that were calibrated for non-evaporating main-sequence stars \citep{VerbuntZwaan1981,Rappaport1983}, and it can reproduce many of the features of the observed black widow population, including novel direct optical observations of the companions themselves \citep{Draghis2019}. A key aspect of this model is the deposition of pulsar irradiation energy in the companion's atmosphere, which slows down its Kelvin--Helmholtz contraction. As a result, black widow companions become substantially inflated compared to main-sequence stars as they lose mass and they retain radiative cores even at masses well below $0.3\, {\rm M_\odot}$. This deviation from the main sequence, as well as mounting theoretical and observational evidence that even fully convective stars can generate strong magnetic fields through a convective dynamo \citep[e.g.][]{ToutPringle1992,Dobler2006,Browning2008,Morin2008,Christensen2009, Knigge2011,El-Badry2022,SarkarTout2022}, implies that magnetic braking may continue to play a major role in black widow evolution even at low companion masses. 

Here, we study the convective dynamos of black widow and redback companions in greater detail. \citetalias{GQ21} used a simple relation \citep{Christensen2009} to relate the companion's magnetic field to its internal luminosity, i.e. the difference between its luminosity and the incident pulsar irradiation. This method becomes inaccurate for strong pulsar irradiation, when the net (internal) luminosity is a very small fraction of the companion's total luminosity. Here we take a different approach and calculate the magnetic field directly from the companion's convective velocities, as inferred by the one-dimensional stellar evolution code \textsc{mesa} \citep{Paxton2011,Paxton2013,Paxton2015,Paxton2018,Paxton2019,Jermyn2023}. This approach also enables us to examine how the dynamo is affected by the changing size of the convective region. \citetalias{GQ21} modelled the companion star as essentially fully convective. However, the relative size of the convective envelope changes with the companion's mass and irradiation as it evolves, up to the point that some stars may become almost fully radiative.

Convective dynamos may be the main drivers of black widow evolution. Our broader goal is to relate these exotic binaries to the magnetism of more common stars, such as cataclysmic variables (CVs), where a donor star fills its Roche lobe around a white dwarf instead of a pulsar, and where the relative role of the convective dynamo is still an open question \citep[e.g.][]{El-Badry2022,SarkarTout2022}. Under the influence of the pulsar's strong irradiation, black widow companions venture out of the main sequence and provide a unique opportunity to systematically probe stellar magnetism and magnetic braking in various regimes. 

The remainder of this paper is organized as follows. In Section \ref{sec:model} we describe our updated computational model. Section \ref{sec:results} presents our results: the transformation of main-sequence stars into spider companions (\ref{sec:tracks}), the nature of the companion's magnetic dynamo (\ref{sec:dynamos}), and the evolution of companions that are close to the end of their main-sequence phase (\ref{sec:evolved}). These evolved mass donors are analogous to the shortest period CVs -- AM Canum Venaticorum (AM CVn) stars. We discuss our findings in Section \ref{sec:discussion}.

\section{Computational model}\label{sec:model}

In this section we review the ingredients of our computational model. Sections \ref{sec:rlo}--\ref{sec:mdot} are similar to \citetalias{GQ21} and are briefly repeated here for completeness. The main difference of our model is the calculation of the companion's magnetic field, which we describe in Section \ref{sec:B}.

\subsection{Roche-lobe overflow}\label{sec:rlo}

If evaporation by the pulsar's radiation is too weak to significantly reduce the companion's mass \citep{GinzburgQuataert2020}, then the companion must instead lose mass through Roche-lobe overflow in order to reach the low masses observed in black widow systems. We follow \citetalias{GQ21} and assume that the loss of orbital angular momentum always keeps the companion close to filling its Roche lobe and dictates stable mass loss (we calculate the rate in Section \ref{sec:mdot}). Modelling of black widow optical light curves seems to support this picture for a large portion of the systems \citep{Draghis2019,KandelRomani2023,Mata2023}.

The radius $r$ of a Roche-lobe filling companion is given by
\begin{equation}\label{eq:rlobe}
\frac{r}{a}\simeq 0.49\left(\frac{m}{M}\right)^{1/3},
\end{equation}
where $M\gg m$ is the pulsar's mass and $a$ is the binary separation \citep{Eggleton83}. The orbital period $P$ of the Keplerian orbit is therefore
\begin{equation}\label{eq:period_r}
P=2\upi\left(\frac{r^3}{0.49^3Gm}\right)^{1/2},    
\end{equation}
where $G$ is the gravitational constant. Our $M\gg m$ approximation \citep[typically $M\gtrsim 1.4\, {\rm M}_\odot$; see][]{KandelRomani2023} and Roche-lobe filling assumption allow us to evolve the companion in \textsc{mesa} as a single star, with the orbital period given at any moment by its mean density $\sim mr^{-3}$ using equation \eqref{eq:period_r}.

\subsection{Irradiation}\label{sec:irradiation}

Irradiation of the companion by the pulsar's high-energy photons is implemented in \textsc{mesa} using the $F_\star$--$\Sigma_\star$ method \citep{Paxton2013} by depositing a flux
\begin{equation}\label{eq:F_star}
F_\star=\frac{L_{\rm irr}}{4\upi a^2}    
\end{equation}
at an outer mass column density $\Sigma_\star$, assuming that heat is efficiently redistributed between the tidally locked companion's day and night hemispheres \citepalias[this assumption holds for high-energy photons that are deposited deep enough in the atmosphere; see][]{GQ21}. We choose $\Sigma_\star=180\textrm{ g cm}^{-2}$, given by the pair production cross section, which dominates the interaction of the pulsar's energetic GeV photons with the companion's atmosphere \citep{BetheHeitler34}.

The pulsar's high-energy photon luminosity $L_{\rm irr}$ is one of the free parameters of our model. {\it Fermi} measured $L_{\rm irr}$ for a subset of spider pulsars, finding that it is a significant -- but not constant -- fraction of the pulsar's total spin-down power \citep{Abdo2013}. This subset has been used to test black widow evolutionary models (\citetalias{GQ21}; \citealt{Swihart2022}). 

\subsection{Mass-loss rate}\label{sec:mdot}

Gravitational waves carry orbital angular momentum from the binary on a time-scale of
\begin{equation}\label{eq:t_gw}
    t_{\rm GW}=\frac{5}{32}\frac{c^5a^4}{G^3Mm(M+m)},
\end{equation}
where $c$ is the speed of light \citep{LandauLifshitz1971}. We take $M=1.4\,{\rm M}_{\odot}$ for consistency with the ATNF pulsar catalogue \citep{Manchester2005}, even though many spider pulsars reach masses $M\approx 2\,{\rm M}_\odot$ \citep{Linares2020,KandelRomani2023}.

Gravitational waves carry significant angular momentum only from the shortest period black widows \citepalias[e.g. fig. 1 in][]{GQ21}. Another sink of angular momentum is the companion's wind, which is forced to corotate with its magnetic field up to a large radius -- such that even a low-mass wind can carry significant angular momentum and spin down the star \citep{WeberDavis67}. Black widow companions are close enough to their host pulsars for their spin to be tidally locked to the orbit. 
Therefore, this magnetic braking mechanism ultimately decreases the binary's orbital angular momentum.

Previous studies incorporated an empirical magnetic braking prescription that was calibrated to solitary main-sequence stars \citep{VerbuntZwaan1981,Rappaport1983}. \citet{El-Badry2022} recently demonstrated that this prescription cannot be extrapolated to short-period binaries such as CVs. More crucially, magnetic braking in black widows and redbacks is driven by pulsar ablation, which dominates over the companion's spontaneously emitted wind. For this reason, we adopt a model that was developed from first principles to black widows \citep{GinzburgQuataert2020}. Using a hydrodynamical calculation to evaluate the strength of the ablated wind, the time-scale on which magnetic braking removes orbital angular momentum from a Roche-lobe filling companion is given by
\begin{equation}\label{eq:t_mag}
\frac{t_{\rm mag}}{\textrm{ Gyr}}=2.0\left(\frac{L_{\rm MeV}}{{\rm L}_{\odot}}\right)^{-4/9}\left(\frac{P}{\textrm{h}}\right)^{-34/27}\left(\frac{m}{{\rm M}_{\odot}}\right)^{2/27}\\
\left(\frac{B}{\textrm{kG}}\right)^{-4/3},
\end{equation}
where $B$ is the companion's magnetic field (\citealt{GinzburgQuataert2020}; \citetalias{GQ21}). As explained in \citet{GinzburgQuataert2020}, the ablated wind is launched primarily by photons close to the electron's rest mass ($\sim$ MeV). These photons deposit energy through Compton scattering high enough in the companion's atmosphere, where cooling is inefficient and heating accelerates the material beyond the escape velocity. We denote the pulsar's MeV luminosity by $L_{\rm MeV}$, which is typically unconstrained by observations \citep[{\it Fermi} was sensitive to more energetic GeV $\gamma$ rays from pulsars; see][]{Abdo2013}. We limit this free parameter to $L_{\rm MeV}\lesssim L_{\rm irr}$, such that it is capped by the broader spectrum total irradiation.    

In stable Roche-lobe overflow, the donor star loses its mass on the same time-scale that it loses its angular momentum, up to an uncertain order-unity coefficient which we omit \citep[this depends on the details of the mass transfer; see][]{Rappaport1982}. The companion's mass-loss rate in our model is therefore given by 
\begin{equation}\label{eq:mdot_sum}
\frac{|\dot{m}|}{m}=\frac{1}{t_{\rm GW}}+\frac{1}{t_{\rm mag}},
\end{equation}
with $t_{\rm GW}$ and $t_{\rm mag}$ calculated in equations \eqref{eq:t_gw} and \eqref{eq:t_mag}.

\subsection{Magnetic field}\label{sec:B}

To complete our computational model, we need to evaluate the companion's magnetic field $B$, which together with the irradiation parameters $L_{\rm irr}$ and $L_{\rm MeV}$ uniquely determines the system's evolution. The main improvement of this work compared to \citetalias{GQ21} is a more physically motivated model for $B$, which we describe in this section.

Convective dynamos are thought to generate magnetic fields by a combination of rotation and convection in a conducting fluid. However, the scaling laws that determine the strength of these fields are still in question \citep{Christensen2010,Augustson2019}. We follow \citetalias{GQ21} and assume that the magnetic energy density is in equipartition with the kinetic energy density of convective eddies

\begin{equation}\label{eq:equipartition}
    B^2\sim\rho v_{\rm conv}^2,
\end{equation}
where $\rho$ is the fluid density and $v_{\rm conv}$ is the convective velocity. This scaling is thought to describe the fields of fast rotating convective dynamos with eddy turnover times that are much longer than the spin period \citep{Christensen2009}; this is the regime of tidally locked spider companion stars with periods of several hours. For simplicity, and due to the uncertainties of the dynamo theory, we omit potential order-unity coefficients from equation \eqref{eq:equipartition}. Equation \eqref{eq:t_mag} indicates that our magnetic braking prescription is sensitive only to the combination $L_{\rm MeV}B^3$, such that these order-unity coefficients are absorbed by the parameter $L_{\rm MeV}$. 
The convection carries an energy flux of
\begin{equation}\label{eq:f_conv}
    F_{\rm conv}\sim\rho v_{\rm conv}^3,
\end{equation}
such that the magnetic field scales as \citep{Christensen2009}
\begin{equation}\label{eq:B_Fconv}
    B\sim\rho^{1/6}F_{\rm conv}^{1/3}.
\end{equation}

In fully convective single stars and brown dwarfs, the convective flux can be related to the luminosity that has to be transported outwards $L\sim r^2 F_{\rm conv}$, yielding a simple expression $B\sim m^{1/6}r^{-7/6}L^{1/3}$, where we used the mean density $\rho\sim mr^{-3}$ \citep{Reiners2009}. The luminosity of black widow companions, on the other hand, is dominated by atmospheric re-emission of the pulsar's irradiating flux $L\approx F_\star\upi r^2$, with only a much smaller internal luminosity $L_{\rm int}\ll L$ delivered by convection from the companion's interior. \citetalias{GQ21} accounted for this effect by scaling $B\sim m^{1/6}r^{-7/6}L_{\rm int}^{1/3}$.

Here we take a more direct approach. Instead of scaling $B$ with global quantities ($m$, $r$, and $L_{\rm int}$), we evaluate $B$ locally at each point using equation \eqref{eq:equipartition} with the local $\rho$ and $v_{\rm conv}$ (as computed by \textsc{mesa}). We then average $B$ over the convective envelope
\begin{equation}\label{eq:avgB}
    B=\frac{\int_{v_{\rm conv}>0}{\rho^{1/2}v_{\rm conv}{\rm d}m}}{\int_{v_{\rm conv}>0}{\rm d}m},
\end{equation}
assuming that convection effectively transports $B$. Averaging by radius rather than by mass does not change the results significantly; in any case the field is dominated by the bottom of the convection zone (perhaps it is more physically motivated to average the energy density by volume, yet all averages yield similar results in our case). 

Our method has two major advantages compared to \citetalias{GQ21}. First of all, at high values of $L_{\rm irr}$, the internal luminosity $L_{\rm int}$ is several orders of magnitude lower than the total luminosity $L_{\rm int}\ll L\approx L_{\rm irr}(r/2a)^2$ (see \citetalias{GQ21}), making it numerically challenging to compute accurately. Equation \eqref{eq:avgB} bypasses this technical difficulty by using $v_{\rm conv}$, which is directly computed by \textsc{mesa} from the net flux through each cell.
More crucially, as we show below, strong irradiation inhibits convection and pushes the tachocline outwards, thinning the convective envelope. This effect was missed by \citetalias{GQ21} who assumed that convection always penetrates deep inside the companion star, leading to an overestimate of $B$ for high values of $L_{\rm irr}$. Equation \eqref{eq:avgB} corrects this by integrating only over the convective zone. The new method will thus enable us to adequately describe the dynamos of various black widow and redback companions, which range from fully convective to almost fully radiative structures, depending on $L_{\rm irr}$.  

\section{Results}\label{sec:results}

In Section \ref{sec:tracks} we calculate evolutionary tracks that match the observed distribution of black widow companion masses and orbital periods by fitting our model's free parameters $L_{\rm irr}$ and $L_{\rm MeV}$. In Section \ref{sec:dynamos} we discuss the properties of the companion's magnetic dynamo as it loses mass along these evolutionary tracks. In Section \ref{sec:evolved} we try to explain some of the anomalous spider observations with progenitors that overflow their Roche lobe close to the end of their main-sequence phase. 

\subsection{Evolutionary tracks}\label{sec:tracks}

\begin{figure}
\includegraphics[width=\columnwidth]{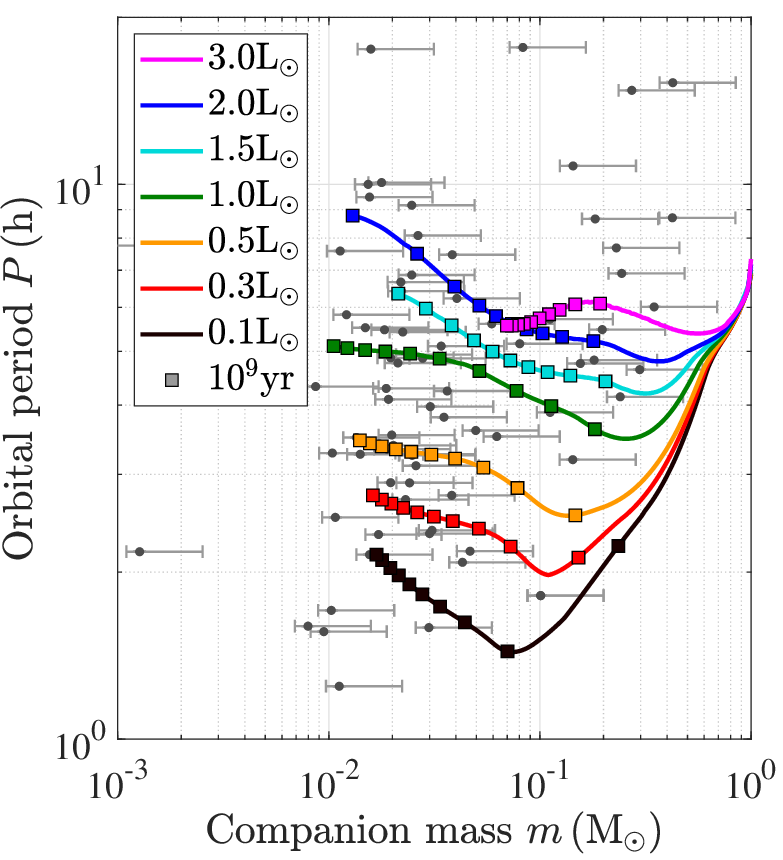}
\caption{Evolutionary tracks of initially $1\,{\rm M}_\odot$ main-sequence companions that lose mass through stable Roche-lobe overflow. The host pulsar emits a high-energy photon luminosity $L_{\rm irr}$ (given in the legend), which slows down the companion's cooling and contraction such that it fills its Roche lobe at a longer orbital period. A fraction of this irradiation -- MeV photons -- ablates a wind off the companion's surface which couples to its magnetic field and removes angular momentum from the binary. This magnetic braking, together with gravitational waves, maintains Roche-lobe overflow and dictates the mass loss rate (see Section \ref{sec:model}). We nominally assume a fraction $f\equiv L_{\rm MeV}/L_{\rm irr}=0.2$, but we choose $f=0.6$ for $L_{\rm irr}=2\,{\rm L}_\odot$ and the maximal $f=1$ for $L_{\rm irr}=3\,{\rm L}_\odot$. Square markers indicate 1 Gyr time intervals (10 Gyr total).}
\label{fig:tracks}
\end{figure}

Fig. \ref{fig:tracks} shows evolutionary tracks of initially $1\,{\rm M}_\odot$ main-sequence companions that are evolved according to our model (Section \ref{sec:model}) with the \textsc{mesa} version r22.11.1. The black widow and redback observations are taken from the ATNF Pulsar Catalogue \url{http://www.atnf.csiro.au/research/pulsar/psrcat} \citep{Manchester2005}, version 1.70 (May 2023). Specifically, we plot all pulsars with spin periods below 30 ms and companions on orbits shorter than 1 d that are classified as either `ultra light' or `main sequence' in the catalogue (this removes white dwarf and neutron star companions from the sample).\footnote{We added to our sample the black widows J1518+0204C \citep{Pallanca2014} and J1653$-$0158 \citep{Nieder2020}, which were miss-classified as helium white dwarfs in the catalogue.} The circular grey markers indicate median companion masses and the error bars encompass 90 per cent of the possible inclinations (i.e. a minimal inclination angle of $25.8^\circ$).

Unsurprisingly, we find the same range of pulsar irradiation luminosities $L_{\rm irr}\approx 0.1-3\,{\rm L}_\odot$ as \citetalias{GQ21}. These luminosities are required in order to lengthen the Kelvin--Helmholtz cooling and contraction time to several Gyr such that companions remain inflated and fill their Roche lobes at the observed orbital periods. Our fitted values of $L_{\rm MeV}$, on the other hand, are somewhat different from \citetalias{GQ21} because of our more careful calculation of the magnetic field $B$. We nominally fit a constant fraction $f\equiv L_{\rm MeV}/L_{\rm irr}=0.2$ in order for the evolutionary tracks to reach the observed mass cutoff at $m\approx 10^{-2}\,\rm{M}_\odot$ in about 10 Gyr; we interpret this cutoff as due to a maximum system age (the two extremely low-mass companions below the cutoff, J2322$-$2650 and J1719$-$1438 with $m\approx 10^{-3}\,{\rm M}_\odot$, are considered outliers here). However, at the highest values of $L_{\rm irr}$, the magnetic field weakens as we explain below, forcing us to fit a higher $f=0.6$ for $L_{\rm irr}=2\,{\rm L}_\odot$ in order to reach the observed mass cutoff. For $L_{\rm irr}=3\,{\rm L}_\odot$ even the maximal $f=1$ is not enough -- the star evolves too slow -- unable to reproduce the longest period observed black widows.

\subsection{Magnetic dynamos}\label{sec:dynamos}

\begin{figure}
\includegraphics[width=\columnwidth]{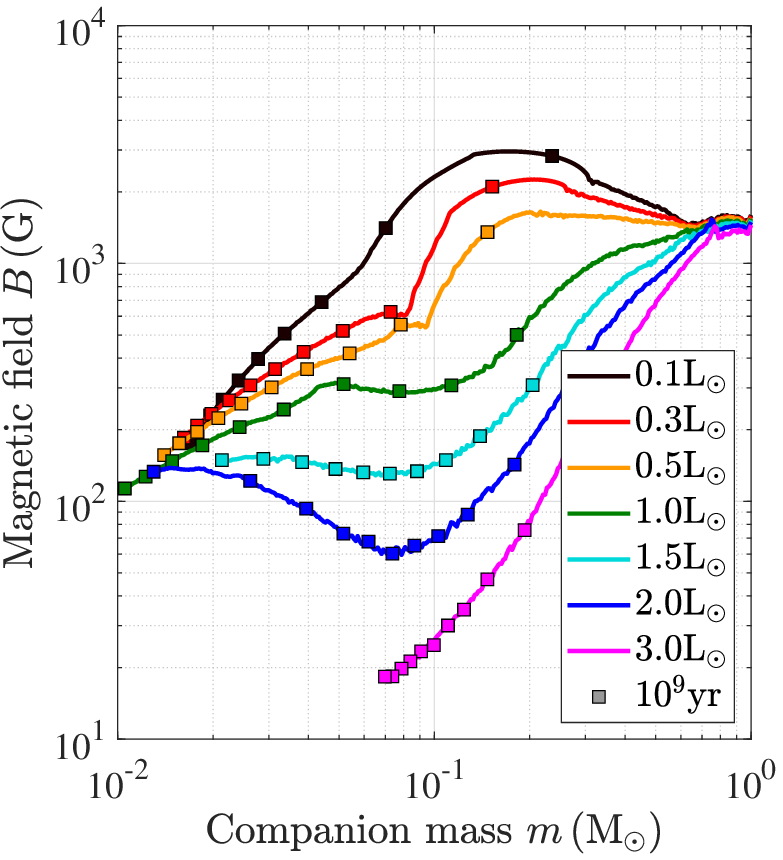}
\caption{The companion's magnetic field $B$ along the tracks presented in Fig. \ref{fig:tracks}. $B$ is calculated by assuming equipartition between the magnetic and kinetic energies in the convective envelope (see Section \ref{sec:B} for details).}
\label{fig:MagB}
\end{figure}

Fig. \ref{fig:MagB} shows our estimate for the companion's magnetic field $B$, assuming equipartition between the magnetic and kinetic energies in the convective envelope (see Section \ref{sec:B}). For most of the tracks, our results are in reasonable agreement with \citetalias{GQ21} -- fields of order $B\sim 10^2-10^3\textrm{ G}$ that weaken with rising $L_{\rm irr}$, because it reduces the internal luminosity $L_{\rm int}$ and with it the convective flux. Both here and in \citetalias{GQ21} the fields converge to similar values at low companion masses (which comprise the majority of spider pulsars).
Interestingly, \citet{Wadiasingh2018} inferred $\sim\,$kG fields from a pressure balance between the companion's magnetosphere and the pulsar wind, which is required to explain the observed intrabinary shock \citep[see also][who derive somewhat weaker fields]{Polzin2018, Miao2023}. 

\begin{figure}
\includegraphics[width=\columnwidth]{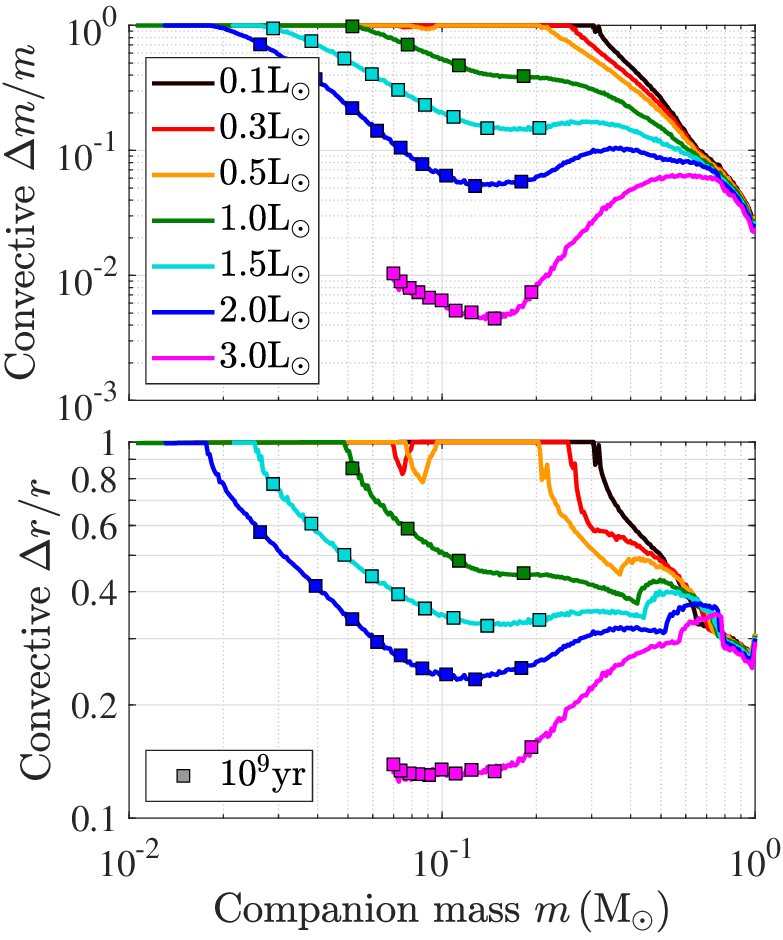}
\caption{The mass (top panel) and radius (bottom panel) fractions of the companion's convective envelope along the tracks presented in Fig. \ref{fig:tracks}. Main-sequence stars become fully convective at $m\approx 0.3\,{\rm M}_\odot$. The pulsar's irradiation $L_{\rm irr}$ (given in the legend) reduces the internal luminosity $L_{\rm int}$ and inhibits convection, such that companions become fully convective at lower masses. For $L_{\rm irr}\gtrsim 3\,{\rm L}_\odot$, the companion becomes almost fully radiative, weakening the magnetic field (Fig. \ref{fig:MagB}). To avoid clutter, the time-interval markers are plotted only before companions become fully convective.}
\label{fig:conv_panels}
\end{figure}

For extremely strong pulsar irradiation $L_{\rm irr}\gtrsim 3\,{\rm L}_\odot$, however, our magnetic fields fall by an order of magnitude below \citetalias{GQ21}. The reason for this can be found in Fig. \ref{fig:conv_panels}, where we show the extent (by mass $\Delta m$ and by radius $\Delta r$) of the companion's convective envelope. The internal luminosity $L_{\rm int}$ is so low in this case that it can be evacuated from the star almost without convection -- the star becomes almost fully radiative, with only a thin convective envelope. The density $\rho$ of this outer shell is much lower than the stellar interior density, leading to a reduction in $B$ for a given convective flux $F_{\rm conv}$, as seen in equation \eqref{eq:B_Fconv}. This effect was missed by \citetalias{GQ21} who assumed that $\rho$ is close to the companion's mean density and thus overestimated $B$. Such high values of $L_{\rm irr}$ have been directly measured for several spider pulsars \citep{Abdo2013,Swihart2022}, necessitating our improved treatment of the convective dynamo.

In addition to its lower density, a geometrically thin convective shell presumably generates a smaller-scale magnetic field, with dominant multipole orders $\sim r/\Delta r\gg 1$, rather than a dipole. It has been argued that such small-scale fields are much less efficient in removing angular momentum \citep{Garraffo2015,Garraffo2016,Reville2015,Garraffo2018}, potentially weakening magnetic braking for high $L_{\rm irr}$ even beyond our model. We do not take this additional effect into account here; our highest $L_{\rm irr}$ models evolve too slow to become black widows in any case.\footnote{\citet{SarkarTout2022} also weaken magnetic braking for thin envelopes using their $\gamma$ parameter, but the physical motivation is different.}

Fig. \ref{fig:conv_panels} demonstrates that spider pulsars can be used to benchmark magnetic dynamo and magnetic braking theories. The structure of their companions changes non-trivially as they lose mass, depending on the value of $L_{\rm irr}$ (which can be mapped to their orbital period, see Fig. \ref{fig:tracks}). Short-period spider companions ($L_{\rm irr}\lesssim 0.5\,{\rm L}_\odot$) resemble main-sequence stars and become fully convective at $m\approx 0.3\,{\rm M}_\odot$. Companions orbiting at intermediate periods ($L_{\rm irr}\approx {\rm L}_\odot$) emit less internal heat because of the irradiation boundary condition and retain their radiative cores down to much lower masses $m<0.1\,{\rm M}_\odot$, as already noticed by \citetalias{GQ21}. At the longest orbital periods ($L_{\rm irr}\gtrsim 2\,{\rm L}_\odot$), the radiative core even penetrates outwards as companions lose mass (at least initially), leaving only a thin convective envelope. 

This variety of spider companion structures, ranging from fully convective to almost fully radiative, depending on their mass and orbital period, may be utilized to test the relative roles of the convective dynamo and of the tachocline in generating magnetic fields \citep[e.g.][]{SarkarTout2022}. Specifically, the companion mass distribution is a direct indicator of the magnetic braking rate, which dictates the mass-loss timescale. In the future, we plan to constrain and calibrate different magnetic braking prescriptions by testing them against this observed distribution. Here, as a first step, we limited ourselves to fitting the specific model of \citet{Christensen2009} and \citetalias{GQ21} in order to demonstrate the potential of these exotic binary systems.

\subsection{Evolved donors}\label{sec:evolved}

Fig. \ref{fig:tracks} shows that there are several observations that cannot be explained by Roche-lobe overflow of main-sequence stars. In this section we demonstrate that most of these observations can be reproduced by the well-known bifurcation of the evolutionary trajectories of evolved donor stars \citep{PylyserSavonije1988,PylyserSavonije1989,Podsiadlowski2002}. Specifically, for stars that overflow their Roche lobe at a time $t_0$ that is close to the end of their main-sequence lifetime $t_{\rm ms}$ there is an interesting interplay between the mass loss and the nuclear burning. As $t_0/t_{\rm ms}$ increases, donors enrich their interiors in helium before losing their envelopes. These dense helium-rich companions evolve to {\it short orbital periods}. Above some critical $t_0\approx t_{\rm ms}$, however, the donor develops a degenerate helium core, while the envelope expands to become a red giant. These low-density giants evolve to {\it long orbital periods} before their envelopes are lost, leaving behind a detached white dwarf (i.e. the degenerate helium core) orbiting the pulsar.  

\subsubsection{Short orbital periods}

The shortest period black widows ($P\lesssim 2\textrm{ h}$) with masses $m\approx 10^{-2}\,{\rm M}_\odot$ lie below all of our nominal tracks. In fact, these observations lie even below the hypothetical $L_{\rm irr}=0$ track, implying donors that are denser than fully degenerate (i.e. non inflated) hydrogen-rich stars (see fig. 1 of \citetalias{GQ21}). By balancing the gravitational pressure $\sim Gm^2r^{-4}$ with the electron degeneracy pressure
\begin{equation}\label{eq:deg_pressure}
    p_e\sim\frac{\hbar^2}{m_e}n_e^{5/3}=\frac{\hbar^2}{m_e}\left(Y_e\frac{\rho}{m_p}\right)^{5/3}, 
\end{equation}
substituting for the mean density $\rho\sim mr^{-3}$,
and assuming Roche-lobe overflow (equation \ref{eq:period_r}), we find that the orbital period of a degenerate donor scales as
\begin{equation}\label{eq:period_evolved}
    P\sim\frac{1}{(G\rho)^{1/2}}\propto\frac{Y_e^{5/2}}{m},
\end{equation}
where $\hbar$ is the reduced Planck constant, $m_e$ and $m_p$ are the electron and proton (and neutron, approximately) masses respectively, $n_e$ is the electron number density, and $Y_e$ is the electron to nucleon (proton and neutron) number fraction. Equation \eqref{eq:period_evolved} indicates that helium-rich donors ($Y_e\to 1/2$) may evolve to minimal orbital periods that are several times shorter than hydrogen-rich donors ($Y_e\to 1$), potentially explaining the shortest period observed black widows.

\begin{figure}
\includegraphics[width=\columnwidth]{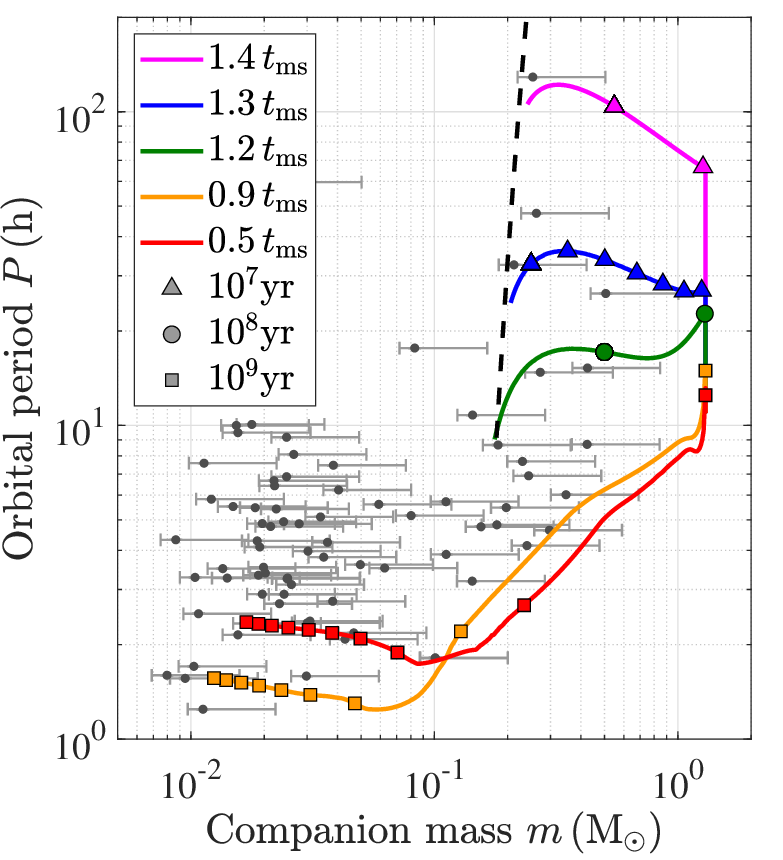}
\caption{Evolutionary tracks of initially $1.3\, {\rm M}_\odot$ companions that overflow their Roche lobe at a time $t_0$ (given in the legend) that is close to the main-sequence lifetime of a solitary star $t_{\rm ms}=2.7\textrm{ Gyr}$. The pulsar's irradiation is $L_{\rm irr}=0.3\,{\rm L}_\odot$ and $L_{\rm MeV}/L_{\rm irr}=0.2$. The $t_0\lesssim t_{\rm ms}$ tracks develop hydrogen poor (helium rich) interiors and terminate at an age of 10 Gyr. The $t_0 \gtrsim t_{\rm ms}$ tracks evolve into red giants with degenerate helium cores and are terminated when the mass of their hydrogen envelope drops below $0.05\,{\rm M}_\odot$ -- a proxy for Roche-lobe detachment (the more accurate \citealt{TaurisSavonije99} cutoff is given by the dashed black line).}
\label{fig:evolved}
\end{figure}

In Fig. \ref{fig:evolved} we demonstrate that somewhat evolved companions that overflow their Roche lobe close to the end of their main sequence $t_{\rm ms}$ develop helium-rich interiors and naturally evolve to short-period black widows in the framework of our model. Specifically, a donor that initiates mass transfer at $t_0=0.9\,t_{\rm ms}$ (the orange track) leaves behind an $\approx 10^{-2}\,{\rm M}_\odot$ remnant that is enriched to about 90 per cent helium by mass and is thus able to explain the shortest period black widows even when accounting for modest inflation due to the pulsar's irradiation (we choose $L_{\rm irr}=0.3\,{\rm L}_\odot$ as an example). However, only massive progenitors have short enough $t_{\rm ms}$ to develop helium-rich interiors and be reduced to $\sim 10^{-2}\,{\rm M}_\odot$ through stable Roche-lobe overflow within the age of the universe. We therefore choose as an example an initial companion mass of $1.3\,{\rm M}_\odot$, for which there is still a convective envelope that enables magnetic braking and the mass ratio is still low enough such that the mass transfer is stable \citep{Ge2015}. 

A similar evolved-donor channel has been invoked to explain the shortest period CVs (AM CVn stars) and ultra-compact X-ray binaries \citep[][]{Tutukov1985,FedorovaErgma1989, Podsiadlowski2002,Podsiadlowski2003}. In our case, only moderate fine tuning is required for this scenario -- the donor can initiate mass transfer during about one tenth of its main-sequence lifetime, and $L_{\rm irr}$ can be an appreciable fraction of ${\rm L}_\odot$ \citep[as in fact measured for the short-period black widow J1653$-$0158; see][]{Swihart2022} -- and the system still evolves to short orbital periods that are consistent with most of the observations. A more fine tuned scenario (i.e. $t_0$ closer to $t_{\rm ms}$, or lower $L_{\rm irr}$) may explain the recently detected shortest period spider pulsars, with periods of $\approx 1\textrm{ h}$ \citep{Burdge2022,Pan2023}. \citet{Guo2022} proposed an alternative scenario to explain short-period black widows, with helium-burning (rather than main-sequence) stars commencing Roche-lobe overflow after completely losing their hydrogen envelopes.

\subsubsection{Long orbital periods}

Another group of observations beyond the reach of our nominal tracks (Fig. \ref{fig:tracks}) consists of the longest period redbacks -- with $m\sim 0.3\,{\rm M}_\odot$ and $P\gtrsim 7\textrm{ h}$. Extrapolating the results of Fig. \ref{fig:tracks} suggests unrealistically high $L_{\rm irr}$ for these systems \citep[beyond the values measured by][]{Abdo2013}. Moreover, for such high $L_{\rm irr}$, we expect magnetic braking to be highly inefficient (see Section \ref{sec:dynamos}).

In Fig. \ref{fig:evolved} (where we plot all the observations up to an orbital period of 10 d) we demonstrate that the Roche-lobe overflow of more evolved companions ($t_0\gtrsim t_{\rm ms})$ may potentially explain these observations without resorting to high $L_{\rm irr}$. Such evolved companions develop a degenerate helium core, burn hydrogen in a thin shell around the core, while the rest of the hydrogen envelope expands to become a red giant. The low mean densities of these red giants cause them to fill their Roche lobes at very long orbital periods $P$. Quantitatively, the radius of a red giant $r$ is dictated by the mass of its helium core, approximately as $r\propto m_{\rm core}^{9/2}$ \citep{RefsdalWeigert1970,Rappaport1995}. By substituting this relation in equation \eqref{eq:period_r} we find
\begin{equation}\label{eq:giant_rad}
    P\propto\left(\frac{r^3}{m_{\rm core}+m_{\rm env}}\right)^{1/2}\propto\left(\frac{m_{\rm core}^{27/2}}{m_{\rm core}+m_{\rm env}}\right)^{1/2},
\end{equation}
where $m_{\rm env}\equiv m-m_{\rm core}$ is the mass of the hydrogen envelope.

When the mass of the hydrogen envelope drops (because of nuclear burning as well as Roche-lobe overflow) and becomes comparable to the mass of the burning shell $\sim 10^{-2}\,{\rm M}_\odot$, it drastically contracts, and the companion transforms into a compact helium white dwarf (the red giant's degenerate core). During the contraction, the companion recedes within its much larger Roche lobe, leaving behind a detached pulsar--white dwarf binary \citep{RefsdalWeigert1969,RefsdalWeigert1971,Phinney1992}. At the moment of detachment $m_{\rm env}\ll m_{\rm core}\approx m$, such that equation \eqref{eq:giant_rad} predicts a correlation between the white dwarf's mass and its orbital period $P\propto m^{25/4}$ (the period stops evolving shortly after detachment because magnetic braking, tidal locking, and gravitational waves are all very weak for such compact companions on long orbits). We plot a more accurate version of this correlation \citep{TaurisSavonije99} in Fig. \ref{fig:evolved} as a dashed black line -- this is approximately where our evolutionary tracks terminate. Most of the long-period redback systems lie to the right of this line, such that they can be potentially explained by red giant donors that are on their way to form pulsar--white dwarf binaries. 

The idea that some of these long-period redbacks (referred to as the `huntsman' subclass) host giant donors and are the progenitors of typical pulsar--white dwarf binaries has been suggested in the past \citep{Strader2015,Camilo2016,Swihart2017,Swihart2018,Swihart2019}. However, Fig. \ref{fig:evolved} demonstrates that it is not trivial to apply our black widow evolutionary model to giant donors. The analysis of \citet{GinzburgQuataert2020} predicts rapid magnetic braking for giant stars around millisecond pulsars. Specifically, our equation \eqref{eq:t_mag} indicates very short time-scales for stars that fill their Roche lobes at orbital periods of a few days $t_{\rm mag}\sim(P/\textrm{h})^{-34/27}\textrm{ Gyr}\sim 10^7\textrm{ yr}$, as also indicated by the time-interval markers in Fig. \ref{fig:evolved}. Such evolutionary time-scales are even shorter than the red giant's nuclear time-scale ($\sim 0.1\, t_{\rm ms}$, which can be inferred from the pre-mass-loss evolution in Fig. \ref{fig:evolved}), implying that huntsman spiders may be a very short lived phase (which might be inconsistent with their observed occurrence rate), or that our magnetic braking prescription has to be adapted for giant donors.

\section{Discussion}\label{sec:discussion}

We used an ablation-driven magnetic braking mechanism (\citealt{GinzburgQuataert2020}; \citetalias{GQ21}) to explain the formation of spider (black widow and redback) binary pulsars. In this scenario, the pulsar's high-energy radiation ablates a wind off the companion's surface, which couples to the companion's magnetic field and removes orbital angular momentum -- thus maintaining stable Roche-lobe overflow which gradually reduces the companion's mass. Because the ablated wind dominates over the companion's spontaneous outflows, it is inadequate to extrapolate standard magnetic braking prescriptions that were calibrated to non-irradiated stars \citep{VerbuntZwaan1981,Rappaport1983}. Instead, the model requires us to estimate the companion's magnetic field from first principles, which is the focus of this paper.

Here, as a demonstration, we adopted a variation of the \citet{Christensen2009} model for the convective dynamos of fast rotating stars and planets, where the magnetic energy is assumed to be in equipartition with the kinetic energy of convective eddies that carry away internal heat. Unlike previous studies (\citetalias{GQ21}, and in the context of planets, \citealt{YadavThorngren2017}) that related the magnetic field to the star's/planet's global luminosity, we used \textsc{mesa} to compute the magnetic field directly from the convective velocities in the companion's convective envelope.  

Our new method is more suited to study the magnetic fields of black widow companions because it explicitly tracks their convective envelopes. Fig. \ref{fig:conv_panels} shows the variety of possible envelopes as the companions lose mass while being irradiated by different pulsar luminosities $L_{\rm irr}$. High $L_{\rm irr}$ reduces the companion's internal heat transport and inhibits convection, such that companions range from fully convective to almost fully radiative, depending on $L_{\rm irr}$ (which is closely correlated with the orbital period $P$; see Fig. \ref{fig:tracks}) and on their mass $m$. For example, with our simple model for the magnetic field, we find that in strongly irradiated spider companions ($L_{\rm irr}\gtrsim 3\,{\rm L}_\odot$) the convective envelope is pushed outwards to very low densities, weakening magnetic braking by an order of magnitude compared to previous studies -- halting the binary's evolution. 

In the future, our method, the growing sample of observed spider pulsars, and the rich variety of their companions' radiative--convective structures (Fig. \ref{fig:conv_panels}), could be used to test different dynamo theories \citep[i.e. in addition to][]{Christensen2009}. Specifically, black widows may be ideal laboratories to calibrate the dependence of magnetic dynamos and magnetic braking on the width of the convective layer \citep{SarkarTout2022}, the dominant multipole order of the generated magnetic field \citep{Garraffo2018}, and the existence of a radiative--convective boundary \citep{Knigge2011,Lu2023}. Such a calibration could then be applied to analyse the rotation rates of single stars \citep{Newton2016,Newton2018}, as well as the distribution of binary orbits \citep{El-Badry2022}, specifically of CVs \citep{Knigge2011} and low-mass X-ray binaries \citep{Gossage2023}.

In addition to the bulk of spider pulsar observations, which can be explained by initially main-sequence donors (Fig. \ref{fig:tracks}), we applied our model to companions that overflow their Roche lobe close to the end of their main-sequence lifetime $t_{\rm ms}$. The bifurcation of the evolutionary tracks of such evolved donors could potentially explain both the shortest and the longest measured orbital periods (Fig. \ref{fig:evolved}). Stars that begin losing mass at $t_0\lesssim t_{\rm ms}$ are enriched in helium and evolve to short orbital periods -- analogous to AM CVn stars and ultra-compact X-ray binaries. Similarly to those systems, the shortest period black widows could be detectable gravitational-wave sources \citep{Nelemas2004,Chen2020}. An important difference between AM CVn stars and black widows is the pulsar's irradiation $L_{\rm irr}$, which tends to widen the binary orbit (Fig. \ref{fig:tracks}). Despite this effect, we find that evolved black widows can reach close orbits with only moderate fine tuning, specifically with non-negligible $L_{\rm irr}$ that is consistent with the observations. Stars that begin losing mass at $t_0\gtrsim t_{\rm ms}$, on the other hand, develop degenerate helium cores, transform into red giants, and evolve to long orbital periods before losing their envelopes and leaving behind their helium white dwarf cores. Such long-period redbacks may thus be the progenitors of the most common type of millisecond pulsar binaries \citep[that is, with a helium white dwarf companion; see][]{Manchester2017}. We caution that more work has to be done on the magnetic braking of giants around millisecond pulsars in order to properly estimate the lifetime of these systems \citep[see also][]{Chen2021,Deng2021,SoetheKepler2021}. 

In summary, our theoretical model predicts a variety of spider companion structures (Fig. \ref{fig:conv_panels}), and as a result, a large range of companion magnetic fields (Fig. \ref{fig:MagB}) -- which govern the evolution of these exotic binary pulsars through magnetic braking. Thanks to these unique circumstances, and the prospect of directly measuring the companion's magnetic field $B$ \citep{Li2023, Wang2023}, black widows and redbacks could become a Rosetta Stone for stellar magnetism.  

\section*{Acknowledgements}

We thank Arnab Sarkar, Eliot Quataert, Samuel Swihart, and Zorawar Wadiasingh for discussions. We also thank Daniel Blatman for help with installing \textsc{mesa}. This research was partially supported by the Israel Ministry of Innovation, Science, and Technology through grant number No. 1001572596.

\section*{Data Availability}

The data underlying this article will be shared on reasonable request to the corresponding author.



\bibliographystyle{mnras}
\input{dynamos.bbl}




\bsp	
\label{lastpage}
\end{document}